\documentstyle[psfig, twocolumn]{venice97}
\begin{document}

\setlength{\parindent}{0pt}
\setlength{\parskip}{ 10pt plus 1pt minus 1pt}
\setlength{\hoffset}{-1.5truecm}
\setlength{\textwidth}{ 17.1truecm }
\setlength{\columnsep}{1truecm }
\setlength{\columnseprule}{0pt}
\setlength{\headheight}{12pt}
\setlength{\headsep}{20pt}
\pagestyle{veniceheadings}

\title{\bf DISTANCES AND ABSOLUTE AGES OF GALACTIC GLOBULAR CLUSTERS
FROM HIPPARCOS PARALLAXES OF LOCAL SUBDWARFS\thanks{Based on data from 
the Hipparcos satellite and from Asiago and McDonald Observatories}}

\author{{\bf R.G.~Gratton$^1$, F.~Fusi~Pecci$^{2,3}$, E.~Carretta$^2$, G.~
Clementini$^2$,}
{\bf C.E.~Corsi$^4$, M.G.~Lattanzi$^5$} \vspace{2mm} \\
$^1$Osservatorio Astronomico di Padova, Vicolo dell'Osservatorio 5, 
35122 Padova, Italy\\
$^2$Osservatorio Astronomico di Bologna, Via Zamboni 33, 40126 Bologna, Italy\\
$^3$Stazione Astronomica, 09012 Capoterra, Cagliari, Italy\\
$^4$Osservatorio Astronomico di Monte Mario, Via del Parco Mellini 84, 
00136 Roma, Italy\\
$^5$Osservatorio Astronomico di Torino, Strada Osservatorio 20, 10025 Pino 
Torinese, Italy}
\maketitle

\begin{abstract}
High
precision trigonometric parallaxes from the HIPPARCOS satellite and
accurate metal abundances ([Fe/H], [O/Fe], and [$\alpha$/Fe]) from high
resolution spectroscopy for about 30 local subdwarfs have been used 
to derive distances and ages for a carefully
selected sample of nine globular clusters. 
We find that HIPPARCOS parallaxes are smaller than the corresponding 
ground-based measurements leading, to a longer distance scale ($\sim 0.2$ mag)
and to ages $\sim 2.8$ Gyr younger.
The relation between the zero age horizontal branch (ZAHB) absolute magnitude 
and
metallicity for the nine programme clusters is : 
$$ M_V(ZAHB) = (0.22\pm 0.09)({\rm [Fe/H]}+1.5) + (0.48\pm 0.04)$$
The corresponding Large Magellanic Cloud distance modulus is 
$(m-M)_0=18.61\pm 0.07$.\\
The age of the {\it bona fide} old globular clusters (Oosterhoff II and 
Blue Horizontal Branch) based on the absolute magnitude of the turn-off 
is:
$${\rm Age} = {11.8^{+2.1}_{-2.5}} {\rm Gyr}$$
The present age of globular clusters {\bf does no longer conflict} with 
standard inflationary models for the Universe.
\vspace {5pt} \\


 Key~words: Clusters: globulars -- Cosmology -- Stars: basic parameters --
Stars: stellar models {}.

\end{abstract}

\section{INTRODUCTION}

\begin{table*}
\caption{\em Basic data for the field subdwarfs}
  \begin{center}
    \leavevmode
    \footnotesize
\begin{tabular}{rrrlccccccl}
\hline \\[-5pt]
 HIP.&HD/&$\pi$&$\delta \pi/\pi$&$V_0$&$M_v$&$(B-V)_0$& [Fe/H] &
[O/Fe] & [$\alpha$/Fe] & Note \\
 No.& Gliese&(mas)\\
\hline \\[-5pt]
\\
\multicolumn{11}{c}{Stars with Hipparcos parallaxes}\\
\\
 14594&  19445& 25.85&0.044&$8.050\pm 0.010$&$5.11\pm 0.09$&
$0.454\pm 0.018$&$-1.91\pm 0.07$& 0.56 & 0.38 &    \\
 15797&G078-33& 39.10&0.032&$8.971\pm 0.009$&$6.93\pm 0.07$&
$0.982\pm 0.002$&$-0.41\pm 0.07$&      & 0.16 &    \\
 66509& 118659& 18.98&0.064&$8.820\pm 0.010$&$5.20\pm 0.14$&
$0.674\pm 0.002$&$-0.55\pm 0.07$& 0.51 & 0.08 &    \\
 72998& 131653& 20.29&0.074&$9.512\pm 0.002$&$6.05\pm 0.16$&
$0.720\pm 0.000$&$-0.63\pm 0.07$& 0.36 & 0.31 &    \\
 74234& 134440& 33.68&0.050&$9.441\pm 0.001$&$7.08\pm 0.11$&
$0.853\pm 0.000$&$-1.28\pm 0.07$&      & 0.15 &    \\
 74235& 134439& 34.14&0.040&$9.073\pm 0.002$&$6.74\pm 0.08$&
$0.773\pm 0.000$&$-1.30\pm 0.07$&      & 0.29 &    \\
 78775& 144579& 69.61&0.008&$6.660\pm 0.000$&$5.87\pm 0.02$&
$0.734\pm 0.004$&$-0.52\pm 0.13$&      &      &    \\
 95727& 231510& 24.85&0.062&$9.004\pm 0.003$&$5.98\pm 0.13$&
$0.782\pm 0.002$&$-0.44\pm 0.07$& 0.34 & 0.14 &    \\
100568& 193901& 22.88&0.054&$8.652\pm 0.002$&$5.45\pm 0.11$&
$0.555\pm 0.003$&$-1.00\pm 0.07$& 0.35 &      &    \\
112811& 216179& 16.66&0.086&$9.333\pm 0.003$&$5.44\pm 0.18$&
$0.684\pm 0.002$&$-0.66\pm 0.07$& 0.45 & 0.29 &    \\
\\
\multicolumn{11}{c}{Stars in Reid's list}\\
\\
 57450&G176-53& 13.61 &0.113&$ 9.92\pm 0.03$&$5.47\pm 0.25$&
$0.55\pm 0.01$&$-1.26\pm 0.07$&        &      &    \\
103269&G212-07& 14.24 &0.103&$10.18\pm 0.06$&$5.85\pm 0.22$&
$0.59\pm 0.02$&$-1.48\pm 0.16$&        &      &E(B-V)=0.03\\
106924&G231-52& 15.20 &0.080&$10.19\pm 0.06$&$6.04\pm 0.17$&
$0.58\pm 0.02$&$-1.60\pm 0.16$&        &      &E(B-V)=0.05\\
\\
\multicolumn{11}{c}{Stars with good ground-based parallaxes}\\
\\
 18915&  25329& 53.7 &0.026&$8.506\pm 0.001$&$7.15\pm 0.06$&
$0.863\pm 0.003$&$-1.69\pm 0.07$&      &      &    \\
 79537& 145417& 71.1 &0.090&$7.531\pm 0.001$&$6.74\pm 0.19$&
$0.815\pm 0.006$&$-1.15\pm 0.13$&      &      &    \\
\hline \\[-5pt]
\end{tabular}
  \end{center}
\end{table*}
Globular Clusters (GC) are among the 
oldest objects in our Galaxy, their age
thus provides a very stringent lower limit to the age $t$\ of the Universe. 
Recent determinations suggest 
values 
of about $14-16$ Gyr 
for the age of globular clusters  (t=$15.8\pm 2.1$~Gyr, 
 \cite{bh95}; t=$14.6\pm 2.5$~Gyr  \cite{Cha96};
t=$15^{+5}_{-3}$~Gyr,  \cite{Vdb96}). 
These values are in conflict 
with the age of the Universe derived from the most recent estimates 
of the Hubble constant and the standard cosmological Eistein-de
Sitter model. Since, most recent 
determinations of $H_0$ are in the range of $55-75$  
km s$^{-1}$Mpc$^{-1}$\, ($H_0=73 \pm 10$ km/s/Mpc,  \cite{Fre97};
$H_0=63.1 \pm 3.4 \pm2.9$ km/s/Mpc,  \cite{Ham96}; 
$H_0=58 \pm 7$ km/s/Mpc,  \cite{Sah97}, $H_0=56 \pm 7$ km/s/Mpc, 
\cite{st97}),   
the age of the Universe is constrained to be $t<11.6$~Gyr in an
Einstein-de Sitter model, and $t<14.9$~Gyr in a flat Universe with
$\Omega_m=0.2$. 
Current ages for globular clusters result then  
uncomfortably large in the framework of standard cosmological models. 

The most robust indicator of the age of globular clusters is 
the absolute visual magnitude of the main sequence turn-off, $M_V(TO)$.
By comparing $M_V(TO)$ with theoretical isochrones of appropriate 
metallicity and helium abundance 
one gets the age of the cluster
(see Eq. 3 by \cite{Ren91}).
Since the directly observable quantity is the apparent magnitude of the 
turn-off, $V(TO)$, it is necessary to know the distance of the cluster.
Indeed, the main difficulty in the derivation of the ages of globular clusters is the
large uncertainty in the distance scale (\cite{Ren91}). 
Independent of the method used, the present ``actual'' accuracy  in GC 
distance modulus determinations is of the order of 0.2 {\rm mag} which,
in turn, corresponds to an uncertainty of about 3 {\rm Gyr} in the age. 

The simplest technique to derive distances to clusters is
to compare their main sequence (MS) with a suitable template (\cite{San70}) :
 i.e.\ an adequate sample of  metal poor non-evolved subdwarfs with
known distances.
Unfortunately, the template main sequence for metal-poor globular clusters has been
up to now quite uncertain due to the paucity of 
metal poor subdwarfs for which reliable data are 
available, i.e.\ absolute magnitudes and colours accurate to better 
than a few hundreds of a magnitude 
(it should be reminded that an error $\Delta V$ = 0.07 mag in the magnitude of 
the turn-off and/or an error $\Delta(B-V)$ = 0.01 mag in its colour,
both translate into an uncertainty of about 1 Gyr in the derived age),
{\it and} metallicity--[Fe/H] known to 
$\sim 0.1$ dex. 
Table 2 of \cite*{Ren91} very clearly illustrates the contributions 
to the total error budget in the age estimate due to the various quantities 
entering the main sequence fitting technique. Uncertainties in distance moduli
are by far the most relevant contributors to the total error affecting the age.

\section{BASIC DATA FOR THE MAIN SEQUENCE FITTING}
\subsection{Subdwarfs Data}

HIPPARCOS has provided absolute parallaxes for over 118,000 stars, with typical
accuracies of $\sim 1$ mas. We have
parallaxes for 99 subdwarfs with metallicities in the 
range $-2.5 <$ [Fe/H] $< 0.2$; this sample was complemented with
data for about 50 stars (mostly metal-rich) having good ground-based
parallaxes and with objects from  \cite*{Rei97} list.\\
 V magnitudes and colours (Johnson $B-V$\ and $V-K$, and Str\"omgren
$b-y$, $m_1$\ and $c_1$)
for the programme stars were obtained from a careful average of the 
data available in the literature. We also used the $V$\ magnitudes and
$B-V$\ colours provided by the Tycho mission (\cite{Gro95}), after
correcting them for the systematic difference with ground-based data 
(0.003~mag in $B-V$). 

High dispersion spectra were acquired for about two thirds of the subdwarfs 
observed by Hipparcos, using 
the 2.7
m telescope at McDonald and the 1.8 m telescope at Cima Ekar (Asiago), and 
were used to derive accurate metal abundances.
The abundance derivation followed the same precepts of the reanalysis of $\sim
300$\ field and $\sim 150$\ GC stars described in \cite*{Gra97} and 
\cite*{cg97}. We 
found that O and the other $\alpha-$elements are overabundant in all
metal-poor stars in our sample. The average overabundances  
in stars with [Fe/H]$<-0.5$ are: 
$$ {\rm [O/Fe]}= 0.38\pm 0.13$$
$$ {\rm [\alpha/Fe]}= 0.26\pm 0.08,$$

(See poster by Gratton et~al. at this same Meeting for a more detailed description of 
the abundance analysis results).

Only 
{\it bona fide} single stars with $\Delta \pi/\pi < 0.12$ and 
$M_V> 5.5$ were used in our age derivation, these are listed in 
Table 1 together with their relevant quantities. 
Stars brighter than this limit were discarded since they may be evolved 
off the 
Zero Age Main Sequence (ZAMS), as well as we eliminated all detected or 
suspected binaries, except the objects were the separation among the two
components is so wide to not disturb the derived absolute magnitudes 
$M_V$ and
the colours.
Errors in the derived $M_V$
are $\leq 0.25$ mag. An accurate analysis 
via MonteCarlo simulations,  
of the 
Lutz-Kelker corrections most appropriate to our sample, 
revealed them to be very small ($\Delta M_{LK}=-0.002$) and thus they
were neglected.
A not negligible source of systematic errors is, 
instead, the 
contamination of the sample of subdwarfs used in the MS 
fitting technique by unresolved binaries.
A very careful procedure was applied to clean up the sample from binaries.
Further, a statistical approach was devised to evaluate systematic corrections 
of our distance moduli for the possible presence of residual undetected 
binaries.

The field subdwarfs listed in Table 1 were divided into metallicity bins 
and used to define template main sequences of the proper metallicity to 
be compared with the 
globular cluster main loci.
To account for the difference between cluster
and field star metallicity, the colours of the subdwarfs were corrected
for the corresponding shift of the main sequences. 

It should be noted that the HIPPARCOS parallaxes 
are
systematically smaller than those listed in the 1991 version of the Yale
Trigonometric Parallax Catalogue (\cite{vaa91}), 
so that the derived absolute magnitudes are on average brighter.
We anticipate that {\bf everything else being constant, globular cluster ages
derived exploiting this new distance scale are about 2.8~Gyrs younger than
those derived from ground-based parallaxes for the local subdwarfs}.

\subsection{Cluster Data}

Main sequence loci for the programme clusters were taken from literature
data. We generally relied upon the quality of the
original photometric data, although such a good quality is not always 
an obvious issue. 
Column 6 of Table 2 gives references for the adopted colour magnitude 
diagrams.\\
Cluster metal abundances have recently been obtained by
\cite*{cg97}, using high dispersion spectra of cluster giants 
and an abundance procedure totally consistent with that used for the
subdwarf sample. 
The availability of 
high quality abundances (standard errors $\sim 0.07$~dex) for both field
subdwarfs and GC giants on a consistent abundance scale is one of the basic 
ingredients of our study.\\
Interstellar reddenings for some of the considered clusters may be uncertain 
by as much as $\sim 0.05$~mag, implying
errors as large as 5~Gyr in the derived ages. 
Reddening values for the programme clusters have been published by  
 \cite*{Zin80} and  \cite*{Ree88}. These values were averaged with
new reddening estimates based on Str\"omgren photometry 
of nearby field stars, to derive the final adopted reddenings listed 
in Column 4 of Table~2. 

\section{GLOBULAR CLUSTERS DISTANCES  VIA MAIN SEQUENCE FITTING}

\begin{figure*} [!ht]
\centerline{\psfig{file=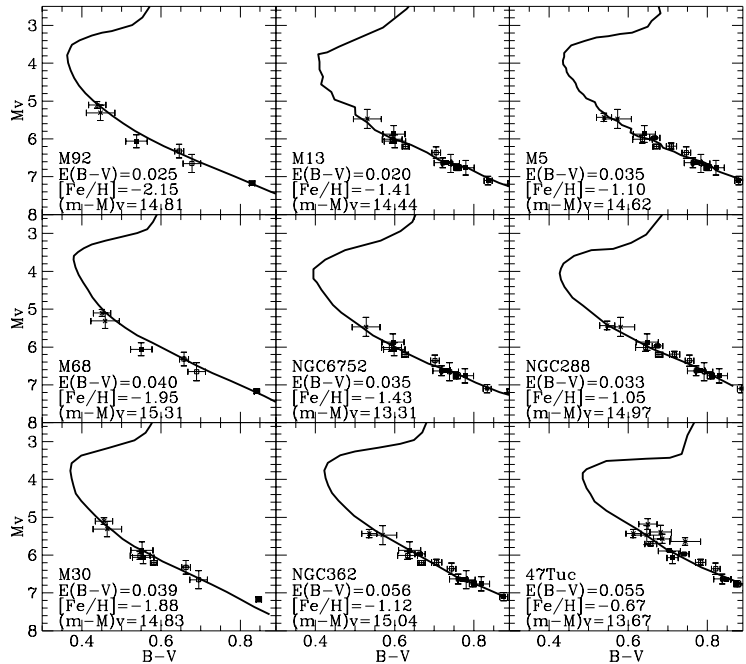,width=10.2cm}}  
\caption{ \em Fits of the fiducial mean loci of the GCs 
with the position of the local subdwarfs. Only {\it bona fide} single
stars with $M_V>5.5$\ (solid squares) were used in the fit. 
Binaries (open squares), and single stars with 
$5<M_V<5.5$ (crosses), are also plotted, but 
they were not used 
in the fit. Each star in the plot was shifted by 0.17 $\times$ {\it p$_i$} 
mag, where 0.17 is the average binary correction, and {\it p$_i$} is the 
probability of star {\it i} to be a binary and ranges from 1 for known 
binaries to
0.08 for supposed single stars.}
\label{f:fig1} 
\end{figure*}

Once template main sequences for the appropriate metallicity are determined,
and cluster reddenings and metallicities are known, cluster distance moduli 
are derived by
least square fitting 
of the apparent magnitude of the cluster main sequence at a
given colour and the absolute magnitude of the template main sequence at the
same 
colour. 
Figure~\ref{f:fig1} displays the fits of the individual GC main sequences 
with
the nearby subdwarfs of the proper metallicity. The derived distance moduli 
are listed in Column 5 of Table~2.
\subsection{Absolute magnitude of the Horizontal Branch}

\begin{table*}
\caption{\em Distance moduli and ages for the programme globular clusters}
  \label{tab:table}
  \begin{center}
    \leavevmode
    \footnotesize
\begin{tabular}{cccccccccccccc}
\hline \\[-5pt]
NGC&Other&[Fe/H]& $<E(B-V)>$&$(m-M)_V$&CMD   & $M_V(HB)$&SC97&VdB97&DCM97&DCM97
&B97\\
   &     &      &           &         &source&         & MLT &MLT  &MLT  &CM &MLT\\
   &     &      &           &         &      &     &(Gyr)&(Gyr)&(Gyr)&(Gyr)&(Gyr)\\
\hline \\[-5pt]
6341& M92&$-2.15$&$0.025\pm 0.005$&14.81&1&$0.24\pm 0.10$&13.6&13.9&13.8&13.0&
12.6\\
4590& M68&$-1.95$&$0.040\pm 0.010$&15.32&2&$0.39\pm 0.11$&11.4&11.6&11.4&11.0
&10.3\\
7099& M30&$-1.88$&$0.039\pm 0.001$&14.95&3,4&$0.25\pm 0.13$&11.1&11.3&11.0&10.7
& 10.1\\
6205& M13&$-1.41$&$0.020\pm 0.000$&14.45&5&$0.50\pm 0.17$&12.5&12.3&12.5&11.9&
11.7\\
6752&    &$-1.43$&$0.035\pm 0.005$&13.32&6&$0.43\pm 0.17$&13.0&12.8&12.7&12.4&
12.1\\
 362&    &$-1.12$&$0.056\pm 0.003$&15.06&7&$0.37\pm 0.13$& 9.0& 8.8& 8.7& 8.5&
 8.0\\
5904& M5 &$-1.10$&$0.035\pm 0.005$&14.61&8&$0.50\pm 0.09$&10.8&10.5&10.4&10.0&
 9.9\\
 288&    &$-1.05$&$0.033\pm 0.007$&14.95&9&$0.45\pm 0.13$&11.3&10.9&10.7&10.5&
10.4\\
 104&47Tuc&$-0.67$&$0.055\pm0.007$&13.63&10&$0.47\pm 0.17$&10.8&10.3&9.8&9.7& 
9.9\\
\hline \\[-5pt]
\end{tabular}
\end{center}
\quotation{CMD source : 1. \cite*{sh88} 2. \cite*{McC87}
3. \cite*{Bol87b} 4. \cite*{Ric88} 5. \cite*{Ric86} 6. \cite*{pd86} 
corrected according to \cite*{Vdb90} 7. \cite*{Bol87a} corrected according to 
\cite*{Vdb90} 8. \cite*{Snd96} 9. \cite*{Buo89} 10. 
\cite*{Hes87}} 
\end{table*}
The absolute magnitude of the Horizontal Branch, $M_V(HB)$, has often been used as a 
standard candle to derive distances and then ages of globular clusters. 
The correct dependence of $M_V(HB)$\ on [Fe/H] is, however
still rather uncertain.
 We have used our distance moduli, that were determined independently of the 
Horizontal Branch (HB) luminosity, to derive a new estimate of the $M_V(HB)$\ {\it vs} [Fe/H] 
relation.
\cite*{Buo89} 
list $M_V(HB)$ values for all the clusters in our sample.
A close inspection of their data revealed that 
these values correspond to a mean magnitude of the HB.
They were transformed to $M_V(ZAHB)$'s using the
relation given by \cite*{San93}, and then combined 
with the metallicities listed in Table~2 to derive the following 
relation between
absolute magnitude of the HB and metallicity : 

$$ M_V(ZAHB) = (0.22\pm 0.09)({\rm [Fe/H]}+1.5) + (0.48\pm 0.04)$$

In Figure~2 we compare our new $M_V(HB)$\ {\it vs} [Fe/H] relation 
with
the predictions of some recent Horizontal Branch models.

The use of the above relationship has a direct impact on several astronomical
issues. For instance, since our value of $M_V$\ for field RR Lyraes at
[Fe/H]$=-1.9$ ($M_V=+0.33\pm 0.07$) is 0.11 mag brighter than the value quoted
by \cite*{Wal92}, we derive a distance modulus for the LMC of
$(m-M)=18.61\pm 0.07$\ (where the error is the statistical one at
[Fe/H]$=-1.9$). If this distance to the LMC is used (rather than the one
frequently adopted from Cepheids: $(m-M)=18.50\pm 0.10$), the extragalactic
distance scale increases (and estimates of the Hubble constant
decrease) by 5 per~cent (for instance, the value of $H_o$\ derived from SN~Ia by 
\cite{Ham96}, would change from $63.1\pm 3.4\pm 2.9$\ to $59.9\pm 3.2\pm 2.8$).
We only remark that the distance modulus
for the LMC based on the HIPPARCOS calibrations is $(m-M)=18.70\pm 0.10$\ from
Cepheids (\cite{fc97}), and $(m-M)=18.6\pm 0.2$\ from Miras (\cite{val97})
 in excellent agreement with our determination. The most
recent determination from the expanding ring around the SN 1987a is
$(m-M)=18.58\pm 0.03$\ (\cite{Pan97}). 
\begin{figure}[!hb] 
\centerline{\hbox{\psfig{figure=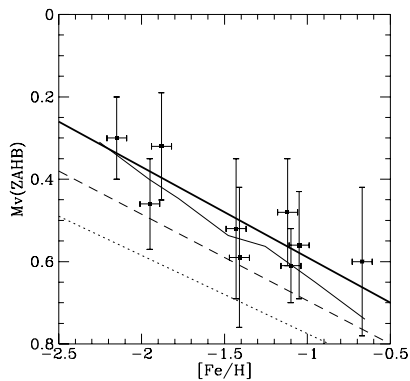,width=6.8cm,clip=}}}  
\caption{\em Runs of the $M_V$(ZAHB) against [Fe/H] for the programme
clusters using our distance moduli and $M_V(HB)$\ from 
Buonanno et~al. (1989) corrected to the ZAHB using Sandage (1993) relation. 
The solid thick line is the weighted least square fit line through the points. 
Also shown are the predictions based on the HB 
models of Caloi et~al. 1997 (solid thin line), VandenBerg 1997 (dotted line), 
and  Salaris et~al. 1997 (dashed line).}
\label{f:fig6} 
\end{figure}  

\section{GLOBULAR CLUSTER AGES}

Ages for the programme clusters were derived from the absolute magnitude of the
turn-off point, $M_V(TO)$. We used the distance moduli of Table~2 and the $TO$\
apparent magnitudes, $V(TO)$\, listed by  \cite*{Buo89}.
Typical errors quoted for
$V(TO)$\ are $0.05-0.10$ leading to errors of $0.09 - 0.13$~mag, and to a 
corresponding average random uncertainty of 12 per~cent in ages. 

Columns 8 to 12 of Table~2 list the ages we derived from
different sets of isochrones. 
Original values from different isochrones have 
been corrected to a common 
value $M_{V_\odot}=4.82\pm 0.02$\
(\cite{Hay85}) for the solar absolute magnitude.
In the isochrones labeled as MLT, convection has
been modeled using the Mixing Length Theory, while those labeled as CM use the
\cite*{cm91} theory. On any other respect, these isochrone sets are
quite similar to each other: they use updated equation of state (including
Debye screening), opacities from the Los Alamos group, and colour
transformations according to \cite*{Kur93}.  
Our data seem to show that some scatter exists in the ages of Oosterhoff II clusters 
(M92, M68 and M30), 
however this scatter is not significantly larger than the expected error bar.
Furthermore, the close similarity of the c-m diagrams for the metal-poor
clusters strongly supports a common age for these clusters. We have therefore 
concluded that the Oosterhoff II clusters 
are indeed
coeval, and that the scatter is due to observational errors. A similar
conclusion is reached for the Blue Horizontal Branch (BHB) clusters 
(M13, NGC288, NGC6752).
Our age estimates for the Oosterhoff I
clusters (M5 and NGC362) are instead $\sim 2.4\pm 1.3$~Gyr lower than
those for the BHB clusters which in turn are similar to those for the 
Oosterhoff II clusters. A rather low age is found also for 47~Tuc. The 
reality of this
age-difference is argument of hot debate. However we must stress that 
the absolute luminosity of the
turn-off point is not the best procedure 
for the derivation of {\it relative} ages, (\cite{Ste96}).
Given the small numbers, the statistical error
bars are not very significant. It seems then wiser to exclude the Oosterhoff I
clusters and 47~Tuc from our estimate of the age of the oldest globular
clusters in our sample. We thus identify the group of {\it bona fide} old
clusters with the Oosterhoff II and the BHB clusters. 

Average ages for our {\it oldest} globular clusters obtained with different
isochrone sets are given in Table~3. Although we cannot exclude the possibility 
that some clusters are older than
the average, this is not required by our observations. We will then assume that
these average values are the best guess for the age of the oldest globular
clusters. 
\noindent
If we use isochrones based on the MLT-convection, the mean age for the six
{\it bona fide} old clusters is $11.8\pm 0.6 \pm 0.4$~Gyr, where the first
error bar is the standard deviation of the mean values obtained for different
clusters, and the second error bar is the spread of ages derived from different
isochrone sets. However, to better quantify the error bars, we
used a MonteCarlo procedure to derive the distribution of total errors and 
to provide the statistical
interval of confidence (95 per~cent range). The following sources of
uncertainty were taken into account : internal errors,  uncertainties in the
solar $M_V$, uncertainties due to the statistical correction for binaries,
reddening scale, metal abundance scale, stellar model code, convection
mechanism, He-sedimentation.
In summary we found that the 
age of the {\it bona fide} old globular clusters (Oosterhoff II and 
Blue Horizontal Branch) based on the absolute magnitude of the turn-off 
is:
$${\rm Age} = {11.8^{+2.1}_{-2.5}} {\rm Gyr}$$

\begin{table}
\caption{\em Mean age for {\it bona fide} old globular clusters}
  \begin{center}
    \leavevmode
    \footnotesize
\begin{tabular}{lc}
\hline \\[-5pt]
Isochrone set                       & Age  \\
                                    &          \\
\hline \\[-5pt]
\multicolumn{2}{l}{Mixing Length Theory}\\
D'Antona, Caloi \& Mazzitelli, 1997(DCM97)&$12.0\pm 0.5$\\
Straniero \& Chieffi, 1997(SC97)         &$12.2\pm 0.4$\\
VandenBerg, 1997(VdB97)                   &$12.0\pm 0.5$\\
Bertelli et al., 1997(B97)              &$11.3\pm 0.4$\\
\multicolumn{2}{l}{Canuto-Mazzitelli Theory}\\
D'Antona, Caloi \& Mazzitelli, 1997(DCM97)&$11.6\pm 0.4$\\
\hline \\[-5pt]
\end{tabular}
  \end{center}
\end{table}

\subsection{Preliminary comparison with other GC ages 
presented at this Meeting}

Pont et~al. have presented at this Meeting results on the age of M92 
based on MS fitting and on a slightly different data base, 
reaching the conclusion that $t_{M92}=14 \pm 1.2$ Gyr.
A few comments on their result is worth here.

The difference in the age derived for M92 cannot be ascribed 
to differences in the data base. Indeed,
if we eliminate from Pont's et~al. data stars at $M_V>5.5$, that may be evolved off the ZAMS, 
and all the detected or suspected binaries, the two data samples 
differ by only one object.

Stars brighter than $M_V=5.5$ may lead
to systematic errors in the derived distance moduli, as it is not 
clear whether GCs and metal-poor field stars are exactly co-eval. For example, 
a 4 Gyr age difference between a calibrating subdwarf at $M_V=5$ and a GC
would lead to a systematic error of 0.05 mag in the distance modulus. 

If binaries are to be included in the sample, 
they must be corrected to account for the contribution due to the secondary 
components. However, the correction is uncertain and strongly depends on the
luminosity function assumed for the secondary components.
We estimate an average binary 
correction of 0.17 mag (based on the luminosity function of Population I 
field stars, \cite{Kro93}), and multiply it by the probability of each 
individual star to be a binary. Pont et~al. use, instead,
an average correction of 0.375 mag, 
based on the binary mass distribution in Praesepe. This is a dynamically 
evolved open cluster, where binaries of higher mass ratio are likely to be 
evaporated and, conversely, equal mass binaries are likely concentrated 
in the center (Kroupa et~al.) thus leading to an overestimate of their number 
and, in turn, of the derived correction.

There also are differences in the adopted metallicity scale and in the 
reddenings, but 
they do not account for the discrepancy in the derived age.

Indeed, the major sources of controversy are the corrections 
that must be applied to the data. They strongly depend on the
selection criteria adopted to build up the sample.
Pont's et~al. data base is formed by two sub-samples, a first one 
(about 60 per~cent of the stars) 
selected before 
the Hipparcos parallaxes were known, and a second one (the remaining 
40 per~cent) selected {\it a posteriori} once parallaxes and colours were known.
Different corrections should thus be applied to the two different 
subsets.
Pont et~al. apply mean bias corrections of +0.064 and $-$0.115 mag
to the subdwarfs and the subgiants, respectively, used to fit the M92 locus. 
Beside distorting the shape of the template sequence, 
these large and uncertain 
corrections 
have a strong impact on the measure of the distance modulus.
Since we do not know in detail how their sample has actually been 
selected, nor if average or individual star corrections have been 
computed and applied, we cannot assess the reliability of their 
derived distance modulus.

Finally, we caution that the use of just one cluster to infer the 
age of the 
{\it oldest} globular clusters may be 
very dangerous since it strongly relies upon the accuracy of that cluster 
photometry. For instance, 
there is a 0.04 mag difference between the colour of the M92 main sequence
as presented by \cite*{hc91} and that by \cite*{sh88}: this 
would translate into a 3 to 4 Gyr difference in the age derived for this
important cluster. 
The use of a not too restricted  sample of 
carefully selected clusters (as we did) reduces the effect of photometric 
errors and  CMD peculiarities.

\subsection{Cosmology}

Assuming a minimum delay of
0.5 Gyrs from the birth of the Universe before the formation of globular
clusters our age estimate is compatible with an Einstein-de Sitter model
if $H_0\leq 64$ km s$^{-1}$Mpc$^{-1}$, and $H_0\leq 83$ km s$^{-1}$Mpc$^{-1}$\
in a flat Universe with $\Omega_m=0.2$. Within the framework of inflationary
models (even in the restricted but more elegant solution of the Einstein-de
Sitter Universe), the presently determined age for the globular clusters is
then consistent with current estimates of the Hubble constant, even without the
$\sim 5$ per~cent reduction which is given by the adoption of the present distance
scale, or that proposed by \cite*{fc97}. We conclude that {\bf at
the present level of accuracy of globular cluster ages, there is no discrepancy
with standard inflationary models for the Universe}.


\begin{thebibliography}{}

\bibitem[\protect\astroncite{Bolte \& Hogan}{1995}]{bh95}
Bolte, M., Hogan C. J. 1995, Nature, 376, 399

\bibitem[\protect\astroncite{Chaboyer et~al.}{1996}]{Cha96}
Chaboyer, B., Demarque, P., Kernan, P.J., Krauss, L.M. 1996,
  Science, 271, 957 

\bibitem[\protect\astroncite{Bertelli et~al.}{1997}]{Ber97}
Bertelli, P., Girardi, L., Bressan, A., Chiosi, C., Nasi, E. 1997,
  in preparation

\bibitem[\protect\astroncite{Bolte}{1987a}]{Bol87a}
Bolte, M., 1987a, ApJ, 315, 469

\bibitem[\protect\astroncite{Bolte}{1987b}]{Bol87b}
Bolte, M., 1987b, ApJ, 319, 760 

\bibitem[\protect\astroncite{Buonanno et~al.}{1989}]{Buo89}
Buonanno, R., Corsi, C. E., Fusi Pecci, F. 1989, A\&A, 216, 80 

\bibitem[\protect\astroncite{Caloi et~al.}{1997}]{Cal97}
Caloi, V., D'Antona, F., Mazzitelli, I. 1997, A\&A, 320, 823

\bibitem[\protect\astroncite{Canuto \& Mazzitelli}{1991}]{cm91}
Canuto V. M., Mazzitelli, I. 1991, ApJ, 370, 5

\bibitem[\protect\astroncite{Carretta \& Gratton}{1997}]{cg97}
Carretta, E., Gratton, R. G. 1997, A\&AS, 121, 95

\bibitem[\protect\astroncite{D'Antona et~al.}{1997}]{Dan97}
D'Antona, F., Caloi, V., Mazzitelli, I. 1997, ApJ, 477, 519

\bibitem[\protect\astroncite{Feast \& Catchpole}{1997}]{fc97}
Feast, M. W., Catchpole, R. M. 1997, MNRAS, 286, L1

\bibitem[\protect\astroncite{Freedman et~al.}{1997}]{Fre97}
Freedman W. L., Madore, B. F., Kennicutt, R. C. 1997, in
  The Extragalactic Distance Scale, eds. M. Livio, M. Donahue, N. Panagia,
  Cambridge Univ. Press, in press

\bibitem[\protect\astroncite{Gratton et~al.}{1997}]{Gra97}
Gratton, R. G., Carretta, E., Castelli, F. 1997a, A\&A, in press

\bibitem[\protect\astroncite{Grossmann et~al.}{1995}]{Gro95}
Grossmann, V., et al. 1995, A\&A 304, 110

\bibitem[\protect\astroncite{Hamuy et~al.}{1996}]{Ham96}
Hamuy et al. 1996, AJ, 112, 2398

\bibitem[\protect\astroncite{Hayes}{1985}]{Hay85}
Hayes, D.S. 1985, IAU Symp. 111, p. 225

\bibitem[\protect\astroncite{Heasley \& Christian}{1991}]{hc91}
Heasley, J. N., Christian, C. A. 1991, AJ, 101, 967 

\bibitem[\protect\astroncite{Hesser et~al.}{1987}]{Hes87}
Hesser, J. E., Harris, W. E., VandenBerg, D. A., Allwright, J. W. 
  B., Shott, P., Stetson, P. 1987, PASP, 99, 739


\bibitem[\protect\astroncite{Kroupa et~al.}{1993}]{Kro93}
Kroupa, P., Tout, C. A., Gilmore, G. 1993, MNRAS, 262, 545


\bibitem[\protect\astroncite{Kurucz}{1993}]{Kur93}
Kurucz, R. L. 1993, CD-ROM 13 and CD-ROM 18

\bibitem[\protect\astroncite{McClure et~al.}{1987}]{McC87}
McClure, R. D., VandenBerg, D. A., Bell, R. A., Hesser, J. E., \&
  Stetson P. B. 1987, AJ, 93, 1144 

\bibitem[\protect\astroncite{Panagia et~al.}{1997}]{Pan97}
Panagia, N., Gilmozzi, R., Kirshner, R.P. 1997, in SN~1987A:
             Ten Years After, eds. M. Phillips and N. Suntzeff, ASP Conf.
             Ser., in press

\bibitem[\protect\astroncite{Penny \& Dickens}{1986}]{pd86}
Penny, A. J., Dickens, R. J. 1986, MNRAS, 220, 845

\bibitem[\protect\astroncite{Reed et~al.}{1988}]{Ree88}
Reed, B. C., Hesser, J. E., Shawl, S. J. 1988, PASP, 100, 545

\bibitem[\protect\astroncite{Reid}{1997}]{Rei97}
Reid, I.N. 1997, AJ, in press

\bibitem[\protect\astroncite{Renzini}{1991}]{Ren91}
Renzini, A. 1991, in Observational Tests of Inflation,
  eds. T. Banday \& T. Shanks, Kluwer, Dordrecht, p. 131

\bibitem[\protect\astroncite{Richer et~al.}{1986}]{Ric86}
Richer, H. B., Fahlman, G. G., 1986, ApJ, 304, 273

\bibitem[\protect\astroncite{Richer et~al.}{1988}]{Ric88}
Richer, H. B., Fahlman, G. G., VandenBerg, D. A. 1988, ApJ, 329, 187 

\bibitem[\protect\astroncite{Saha et~al.}{1997}]{Sah97}
Saha, A., Sandage, A. R., Labhardt, L., Tammann, G. A.,
  Macchetto, F. D., Panagia, N. 1997, ApJ, in press

\bibitem[\protect\astroncite{Salaris et~al.}{1997}]{Slr97}
Salaris, M., Degl'Innocenti, S., Weiss, A. 1997, ApJ, 479, 665

\bibitem[\protect\astroncite{Sandage}{1970}]{San70}
Sandage, A. R., 1970, ApJ, 162, 841

\bibitem[\protect\astroncite{Sandage}{1993}]{San93}
Sandage, A. R., 1993, AJ, 106, 703

\bibitem[\protect\astroncite{Sandage \& Tamman}{1997}]{st97}
Sandage, A. R., Tammann, G.A. 1997, preprint

\bibitem[\protect\astroncite{Sandquist et~al.}{1996}]{Snd96}
Sandquist, E. L., Bolte, M., Stetson, P. B., Hesser, J. E. 1996, ApJ,
  470, 910

\bibitem[\protect\astroncite{Stetson \& Harris}{1988}]{sh88}
Stetson, P. B., Harris, W. E. 1988, AJ, 96, 909 

\bibitem[\protect\astroncite{Stetson et~al.}{1996}]{Ste96}
Stetson, P. B., VandenBerg D.A., Bolte, M. 1996, PASP, 108, 560 

\bibitem[\protect\astroncite{Straniero \& Chieffi}{1997}]{sc97}
Straniero, O., \&  Chieffi, A. 1997, private communication


\bibitem[\protect\astroncite{van Altena et~al.}{1991}]{vaa91}
van Altena, W. F., Lee, J. T., Hoffleit, E. D. 1991, "The General 
  Catalogue of Trigonometric Parallaxes: a preliminary version", Yale
  University electronic preprint

\bibitem[\protect\astroncite{van Leeuwen et~al.}{1997}]{val97}
van Leeuwen, F., Feast M. W., Whitelock, P. A., Yudin, B.
  1997, MNRAS, 287, 955

\bibitem[\protect\astroncite{VandenBerg et~al.}{1990}]{Vdb90}
VandenBerg, D. A., Bolte, M., Stetson, P. B. 1990, AJ, 100, 445

\bibitem[\protect\astroncite{VandenBerg et~al.}{1996}]{Vdb96}
VandenBerg, D. A., Bolte, M., Stetson, P. B. 1996, ARAA, 34, 461 

\bibitem[\protect\astroncite{VandenBerg}{1997}]{Vdb97}
VandenBerg, D. A. 1997, private communication

\bibitem[\protect\astroncite{Walker}{1992}]{Wal92}
Walker, A. R. 1992, ApJ, 390, L81

\bibitem[\protect\astroncite{Zinn}{1980}]{Zin80}
Zinn, R. 1980, ApJS, 42, 19


\end{thebibliography}
\end{document}